\PassOptionsToPackage{sort}{natbib}
\documentclass[twocolumn, trackchanges]{aastex7}

\usepackage{CJKutf8}

\usepackage[caption=false]{subfig}
\usepackage{tabularx}
\usepackage{capt-of}
\usepackage{hhline}
\usepackage{amsmath}	
\usepackage{amssymb}	
\usepackage{textcomp}
\usepackage{upgreek}
\usepackage{listings}
\usepackage{hyperref}
\usepackage{pifont} 

\begin{document}

\title{Observing the First Supermassive Black Holes with JWST}

\author[orcid=0009-0003-1022-6003]{Hyun Song}
\affiliation{Department of Astronomy, University of Texas at Austin, Austin, TX 78712, USA}
\affiliation{Astronomy Program, Department of Physics and Astronomy, Seoul National University, Seoul 08826, Republic of Korea}
\email[show]{thdgus0911@gmail.com}

\author[orcid=0000-0002-4966-7450]{Junehyoung Jeon} 
\affiliation{Department of Astronomy, University of Texas at Austin, Austin, TX 78712, USA}
\affiliation{Cosmic Frontier Center, The University of Texas at Austin, Austin, TX, USA}
\email{junehyoungjeon@utexas.edu}

\author[0000-0003-1282-7454]{Anthony Taylor} 
\affiliation{Department of Astronomy, University of Texas at Austin, Austin, TX 78712, USA}
\affiliation{Cosmic Frontier Center, The University of Texas at Austin, Austin, TX, USA}
\email{anthony.taylor@austin.utexas.edu}

\author[orcid=0000-0003-0212-2979,gname=Volker,sname=Bromm]{Volker Bromm}
\affiliation{Department of Astronomy, University of Texas at Austin, Austin, TX 78712, USA}
\affiliation{Cosmic Frontier Center, The University of Texas at Austin, Austin, TX, USA}
\affiliation{Weinberg Institute for Theoretical Physics, University of Texas at Austin, Austin, TX 78712, USA}
\email{vbromm@astro.as.utexas.edu}

\begin{abstract}
Observing the first generation of supermassive black holes (SMBHs) is essential for understanding their origin and early evolution. The James Webb Space Telescope (JWST) has extended the observational frontier to earlier cosmic epochs than ever before. In this work, we investigate the redshift horizon out to which JWST can detect SMBHs under several representative scenarios. We determine this horizon using two criteria: the minimum SMBH mass required to exceed the JWST sensitivity, and the maximum mass estimated under each seeding and growth scenario. We explore three scenarios for SMBH evolution: (1) a feedback-regulated model, (2) a light seed (LS) scenario with continuous super-Eddington accretion, and (3) a direct-collapse black hole (DCBH) scenario with Eddington-limited accretion. We find that the feedback-regulated model predicts a redshift horizon of $z\sim7$, already exceeded by several observed high-redshift SMBH candidates, but consistent with previous studies suggesting that this model cannot explain early SMBH growth. The LS and DCBH scenarios yield redshift horizons of $z\sim15$ and $z\sim17$, respectively, indicating that JWST may probe SMBHs at substantially earlier times, thus extending the current frontier. We further show that spectroscopic confirmation of SMBHs up to $z\sim16$ may be feasible through [N V] detection with NIRSpec in $\sim100\,\mathrm{hr}$ exposures, or through GN-z11-like spectroscopic diagnostics with longer integration times. Although significant uncertainties remain in the formation and early growth of seed black holes, future SMBH observations at the predicted redshift horizons may serve to distinguish SMBH seeding scenarios. 
\end{abstract}



\section{Introduction}\label{sec:intro}
The James Webb Space Telescope (JWST) has significantly expanded our observational horizon and continues to push the empirical frontier to increasingly high redshifts \citep[e.g.,][]{AdamoRev2025}. Within the standard $\Lambda$CDM model of bottom-up structure formation, there is a limit to how far out we will be able to observe luminous sources due to the disappearance of sufficiently massive host systems at early times \citep[e.g.,][]{Couchman1986,HTL1996,Tegmark1997}. Assessing the redshift horizon of JWST is therefore essential for understanding how far we can directly probe the earliest stages of galaxy formation \citep[e.g.,][]{BrommYoshida2011,Dayal2018}.

In particular, observing the first supermassive black holes (SMBHs) is crucial in revealing their origin and evolution. As luminous quasars with inferred masses of $\sim10^9\,M_\odot$ have been observed at $z\gtrsim6$ \citep[e.g.,][]{2001AJ....122.2850B, 2001AJ....122.2833F, 2011Natur.474..616M}, a key challenge is to understand how they grew so fast within the first billion years \citep[e.g.,][]{Haiman2001}. Their emergence poses a severe
timing problem, most acutely when considering a `light seed' (LS) origin from remnants of metal-free Population~III (Pop III) stars \citep[e.g.,][]{Smith:2019nuc,2020ARA&A..58...27I}.

In line with the known pathway to black hole formation in the local universe, the LS model posits that Pop III~stars collapse into black hole remnants of $\sim10 - 10^2\,M_\odot$ at $z\sim20-30$ \citep[e.g.,][]{1999ApJ...527L...5B, 2001ApJ...551L..27M, 2023ARA&A..61...65K}. However, feedback from the stellar progenitors, including intense photoionization and supernova explosions, expels the surrounding gas, forcing the seeds into a prolonged accretion-starvation phase of roughly $100\,\mathrm{Myr}$ until the surrounding gas recollapses \citep[e.g.,][]{2007MNRAS.374.1557J,2009ApJ...701L.133A, 2014MNRAS.444.3288J}. Moreover, this scenario may require rapid or continuous gas accretion episodes to explain massive quasars observed at high redshift \citep{2015MNRAS.452.1922P}. This difficulty has motivated an alternative origin scenario for SMBHs: heavy seeds, which can alleviate the growth-time problem \citep[e.g.,][]{1995ApJ...443...11E, Woods2019}.

The heavy seed model suggests substantially higher initial BH masses of $10^4$–$10^5\,M_\odot$ \citep[e.g.,][]{Pacucci_MF2017,Becerra2018}. A leading heavy-seed variant is the direct-collapse black hole (DCBH) scenario \citep[e.g.,][]{2003ApJ...596...34B,2006MNRAS.370..289B,Lodato2006}. Such DCBH seed can form within a massive, pristine gas cloud with virial temperature $T_{\mathrm{vir}}\sim10^4\,\mathrm{K}$ that is exposed to strong Lyman–Werner (LW) radiation (11.2–13.6 eV) to dissociate H$_2$ \citep[e.g.,][]{2008MNRAS.391.1961D,Habouzit2016}. In the absence of metals, and by suppressing molecular hydrogen cooling, the gas can undergo monolithic collapse without vigorous fragmentation, forming a supermassive star that rapidly collapses into a heavy seed black hole of $\sim10^5\,M_\odot$ \citep[e.g.,][]{Shibata2002,Reisswig2013}.~\footnote{Heavy seeds could also arise from runaway collisions in dense stellar clusters \citep[e.g.,][]{Zwick_2023,Reinoso_2025}, or from primordial black holes (PBHs; e.g., \citealt{Dayal_PBH_2024,Zhang_PBH_2026}).}

Distinguishing between light and heavy seed scenarios is challenging as the signatures of their initial seeding are erased during subsequent accretion and merger-driven growth, leading to strong degeneracies \citep{2018MNRAS.476..407V, 2018MNRAS.481.3278R,Jeon2025_bhmf,Taylor2025}. However, with the advent of JWST, the redshift horizon for detecting SMBHs has been steadily extended, beginning to probe SMBH evolution in its infancy. Tantalizing SMBH candidates at $z\sim10$ have recently been reported, such as UHZ1 and GN-z11 \citep{Goulding_UHZ12023,2024Natur.627...59M,2024ApJ...960L...1N}, raising expectations that the horizon can be shifted to even earlier epochs \citep[e.g.,][]{2023A&A...677A..88B, 2024ApJ...960L...1N}.

The pursuit of the high-redshift frontier has also been focused on starburst galaxies \citep[e.g.,][]{2019ApJ...883..113D, 2024ApJ...973L..12V,Gandolfi2026}, on supernova explosions \citep[e.g.,][]{Ferrara_PISN2026,Jeon_PISN2026}, and on gamma-ray bursts (GRBs; e.g., \citealt{Salvaterra_GRB_2009,Tanvir_GRB_2009,Toma_GRB_2016}). However, given the energetics and timescales involved, accreting SMBHs clearly are a prime target for the most distant objects accessible to astronomy. We here follow up on earlier studies on the feasibility of observing early SMBHs with JWST \citep[e.g.,][]{2015MNRAS.454.3771P,2017ApJ...838..117N,2020ApJ...897L..16W}, deriving the JWST redshift horizon for different SMBH seeding and growth pathways. We specifically explore whether such differences in redshift horizon can provide a criterion for distinguishing BH seeding models. This exploration may also serve as a heuristic guide for forthcoming surveys that will probe the first black holes to even deeper levels.

The paper is organized as follows. In Section \ref{sec:method}, we introduce the theoretical framework for deriving redshift-dependent minimum and maximum mass limits for observable SMBHs. Section \ref{sec:Results} presents the resulting redshift horizon for different scenarios and prospects for spectroscopic confirmation. Finally, in Section~4 we offer our conclusions and discuss broader implications. We assume ``Planck18" cosmological parameters throughout this investigation \citep{2020A&A...641A...6P}: $H_0 = 67.7{\,}\mathrm{km{\,}s^{-1}{\,}Mpc^{-1}}$, $\Omega_{m} = 0.31$, and $\Omega_{\Lambda} = 0.69$.

\section{Methodology} \label{sec:method}
To determine the redshift limit for SMBH observations, we consider two key conditions on the BH mass: the minimum value required to produce sufficient radiative flux to exceed the JWST sensitivity threshold, and a maximum provided by the abundance constraint that we need to find at least one source in a given JWST survey volume. For the former, we consider luminosities close to the Eddington limit. For the latter, we explore the SMBH abundances corresponding to three representative scenarios: (1) a feedback-regulated model, (2) a LS model with super-Eddington accretion, and (3) a DCBH model with Eddington-limited accretion \citep[e.g.,][]{Jeon2025ApJ...979..127J}.

\subsection{Black Hole Mass Required for JWST Detection} \label{sec:min_mass}

To obtain the lower limit for the SMBH mass, we derive the following threshold condition \citep[e.g.,][]{Bromm_notes2013}:
\begin{equation}
    f_{\rm obs, \nu} = (1+z)\frac{L_{\rm em, \nu}}{4\pi D_L^2} \gtrsim f_{\mathrm{lim}, \nu} \mbox{\ .}
\label{mass_L_ineq}
\end{equation}

Here, $f_{\rm obs, \nu}$ denotes the specific flux observed by the telescope, while $L_{\rm em, \nu}$ is the specific rest frame (emitted) luminosity of the source and $D_L$ is the luminosity distance. We adopt a fiducial JWST NIRCam detection threshold of $f_{\mathrm{lim}, \nu} = 1$\,nJy (AB apparent magnitude 31.4; e.g., \citealt{Gardner:2006JWST}). For simplicity, we assume that the bolometric luminosity of an SMBH is given by its Eddington luminosity, $L_\mathrm{Edd}$, such that we approximately have:
\begin{equation}
    L_{\rm em, \nu} \approx \frac{L_{\rm Edd}}{\nu_{\rm peak}} \mbox{\ ,}
\label{L_emnu}
\end{equation}
where $\nu_{\rm peak}$ is the peak frequency, estimated from Wien’s law for the effective temperature of the SMBH accretion disk. Given its variation with radius, we adopt the maximum effective temperature as a characteristic value for the accretion disk \citep[e.g.,][]{Bonning_2007}, resulting in
\begin{align}
\begin{split}
    \nu_{\rm peak} \approx 
      7\times 10^{16} \,\mathrm{Hz}  \left( \frac{M_{\mathrm{BH}}}{10^6M_\odot}\right)^{-1/4} 
      \mbox{\ .}
\label{nu}
\end{split}
\end{align}
Substituting into Equation~(\ref{L_emnu}) yields
\begin{equation}
    L_{em, \nu} \approx 2\times 10^{27} \,\mathrm{erg~s^{-1}\,Hz^{-1}}  
    \left(\frac{M_{\mathrm{BH}}}{10^6M_\odot}\right)^{5/4}\mbox{\ .}
\label{L_emnu_fin}
\end{equation}

Equation~(\ref{mass_L_ineq}) can then be rewritten to obtain a lower bound for the SMBH mass, as follows:



\begin{equation}
\begin{aligned}
M_{\mathrm{BH}} \gtrsim
10^6 M_\odot\,
\left[\frac{(1+z)}{10} \right]^{4/5}
\left(\frac{f_{\mathrm{lim},\nu}}{1\,\mathrm{nJy}}\right)^{4/5},
\end{aligned}
\label{mass_L}
\end{equation}
where we have assumed that $D_L \propto (1+z)$ at $z\gg 1$.


\subsection{Black Hole Abundance Constraints}\label{sec:abundance}
Considering bottom-up cosmological structure formation \citep[e.g.,][]{Blumenthal_1984,1985ApJ...292..371D}, SMBHs with larger masses are predicted to become increasingly rare at higher redshifts. As mentioned before, this imposes an upper limit on possible SMBH host halo masses, by demanding that we can
find at least one source for a given JWST survey volume. From the abundance constraint, we can in turn derive redshift-dependent upper limits for the SMBH mass for a given growth pathway. 
Throughout this study, we adopt a fiducial survey area of $10{,}000\,\mathrm{arcmin}^2$ (2.78~deg$^2$), which represents a plausible cumulative area achievable by combining multiple current and future JWST surveys \citep[e.g.,][]{2023ApJ...954...31C, eisenstein2023overviewjwstadvanceddeep,finkelstein25}.

\subsubsection{Feedback-Regulated Model} \label{meth:FR}
Using the feedback-regulated model of \citet{10.1093/mnras/stac2182}, based on the energy balance between black hole feedback and host halo gravitational potential energy, we have
\begin{equation}
    M_\mathrm{gas}\frac{GM_\mathrm{h}}{R_\mathrm{VIR}} \simeq f \,c^2M_\mathrm{BH} \mbox{\ ,}
\label{eq:Eequil}
\end{equation}
where the right hand side represents the approximate maximum energy absorbed by the halo gas over time with mass $M_{\rm gas}$, halo mass $M_{\rm h}$, virial radius $R_{\rm VIR}$, $c$ the speed of light, and $f$ denoting the fraction of the SMBH rest mass energy that is deposited in the form of thermal energy. From this relation, we derive the maximum SMBH mass for which the expected number of black holes within the survey volume is of order unity.

Evaluating the Press-Schechter mass function \citep{1974ApJ...187..425P} with the \texttt{COLOSSUS} package \citep{2018ApJS..239...35D}, we can write the number density of halos of mass larger than $M_h$ at a given redshift $z$ as follows:
\begin{equation}
    n(\geq M_\mathrm{h},\,z) = \int_{M_\mathrm{h}}^{\infty} \frac{dn}{dM_\mathrm{h}'} (M_\mathrm{h}',\,z) \, dM_\mathrm{h}'\mbox{\ .}
\label{n_mh}
\end{equation}
Since this function rapidly declines beyond $2\times10^{13}\,M_\odot$ \citep{Shankar_2006}, we here conservatively set the upper limit of the numerical integration to $10^{16}\,M_\odot$, and verified that this choice has negligible impact on our results.

Consequently, the total number of dark‐matter halos with mass exceeding $M_h$ within the survey volume, above a given redshift $z$, is 
\begin{equation}
    N(\geq M_\mathrm{h},\,\geq z) \sim \int_{V(z)}^{V_\infty}  n(\geq M_{\mathrm{h}},\,z') \,dV_\mathrm{com}(z')\mbox{\ ,}
\label{N_mh}
\end{equation}
with a comoving volume element of
\begin{equation}
    dV_{\mathrm{com}}(z) \sim \frac{c}{H_0} \frac{d^2_{\mathrm{com}}(z)}{\sqrt{\Omega_{m}(1+z)^3 + \Omega_{\Lambda}}}\, d\Omega \, dz\mbox{\ .}
\label{V_com}
\end{equation}
where $d_{\mathrm{com}}$ is the comoving distance. Here, the angular integration can be replaced by multiplying with the survey area, $A$, taken as $10{,}000\,\mathrm{arcmin}^2$ throughout this study.

Numerically solving Equation~(\ref{N_mh}), we obtain the maximum halo mass $M_\mathrm{h}$ still satisfying $N(\geq M_\mathrm{h},\,\geq z) \geq 1$, which also sets the maximum limit for the observable SMBH mass according to Equation~(\ref{eq:Eequil}). We choose a technical upper limit of $z=30$ for the numerical integration, again verifying that larger redshift limits do not significantly change the results. Assuming an efficiency factor of $f \sim 0.1$, the relation between black hole and halo mass becomes:
\begin{equation}
    M_\mathrm{BH} \simeq 2\times 10^3\,M_\odot  \left(\frac{1+z}{10}\right) \left(\frac{M_\mathrm{h}}{10^{10}M_\odot}\right)^{5/3} \mbox{\ .}
\label{m_relation}
\end{equation}
For the feedback-regulated scenario, we thus derive the upper bound on the mass of an SMBH that can in principle be observed with JWST, corresponding to an expected number of order unity within the survey area.


\subsubsection{Light Seed Pathway}\label{meth:LS}
\vspace{-0.5em}

We next consider the LS origin with super-Eddington accretion, selecting an initial seed mass that is representative for stellar remnants, as well as an ambitious, yet plausible redshift at which at least one such seed is expected to form within the survey area of $10{,}000\,\mathrm{arcmin}^2$.

The Pop~III initial mass function (IMF) is expected to be top-heavy, based on an extensive body of work \citep[e.g.,][]{Bromm_Review_2013,2023ARA&A..61...65K}. These massive Pop~III stars eventually collapse into black hole remnants with masses of $10-100\,M_\odot$ after a short lifetime of a few million years \citep[e.g.,][]{Heger_2003}. We specifically adopt a light seed mass of $100\,M_\odot$, toward the higher-mass end of the distribution.

Another key parameter is the redshift at which at least one stellar-remnant black hole is expected to form within the survey area. Several studies have suggested that the Pop~III star formation rate density exceeds $10^{-6}\,M_\odot\,\mathrm{Mpc^{-3}\,yr^{-1}}$ at $z \sim 30$ \citep[e.g.,][]{Trenti_2009, 2018MNRAS.479.4544M, Jaacks_2019,Hartwig_2022}. Exploring the maximal growth potential of this scenario, we adopt $z = 30$ as the initial redshift for light seed formation. This choice ensures that Pop~III star-forming minihalos ($\sim$10$^6$ $M_\odot$) are common within the estimated survey volume (assuming $\sim10^4$ arcmin$^2$). Using the GUREFT halo mass function \citep{Yung2024}, we estimate the volume density of minihalos at $z\sim30$ to be $\sim0.3$ Mpc$^{-3}$ dex$^{-1}$. Compared to the survey area, covering a comoving volume of $\sim5\times10^6$ Mpc$^{3}$ at $z\sim30$, numerous minihalos should exist within the survey, forming Pop~III stars and subsequently light BH seeds. However, most of them will remain dormant, and only a few cases will grow at extreme rates \citep[e.g.,][]{Jeon2023,Mehta_2026}. 

Following the formation of light seed black holes, accretion is expected to be temporarily suppressed due to strong outflows. We therefore adopt a fallback time of $100\,\mathrm{Myr}$, motivated by studies of the dynamical, chemical, and thermal evolution of the surrounding gas \citep{2007MNRAS.374.1557J,Jeon_Recovery_2014}. After fallback, we assume that the black hole accretes continuously at twice the Eddington limit, representing an optimistic but commonly adopted growth scenario \citep[e.g.,][]{2014ApJ...784L..38M, 2016MNRAS.458.3047P}. Thus, the maximum mass reached by a seed with a given initial mass and formation redshift can be determined by tracing its subsequent growth. Assuming a constant Eddington ratio of $\lambda_\mathrm{Edd}=2$, the corresponding mass accretion rate can be written as
\begin{equation}
{        \dot{M}_\mathrm{BH} = \lambda_\mathrm{Edd}\frac{L_\mathrm{Edd}}{\eta \, c^2} \mbox{\ ,}
}
\label{edd_accretion}
\end{equation}
with a radiative efficiency $\eta=0.03$–$0.1$, encompassing both radiatively inefficient and efficient accretion flows \citep[e.g.,][]{1974ApJ...191..507T, radeff_2011ApJ...728...98D}. Detailed calculations describing the seed growth are presented in Appendix~\ref{app:growth}.



\subsubsection{DCBH Pathway} \label{meth:DCBH}
Next, we consider the DCBH scenario with accretion at the Eddington limit as an alternative to the case of light seeds under super-Eddington growth discussed above. 

Although the early universe is initially rich in pristine gas, maintaining a sufficiently strong LW radiation field is a demanding requirement to form a DCBH seed. A plausible pathway that satisfies this condition involves DCBH formation in a host system located near a young, star-forming halo, whose intense LW radiation can dissociate $\mathrm{H}_2$ molecules in the neighboring system. In this configuration, the DCBH-forming system can subsequently merge with the neighboring halo, giving rise to an overmassive black hole relative to the host galaxy. The nearby star-forming halo thus not only provides the critical LW flux but can also supply abundant gas, enabling sustained accretion onto the DCBH even after much of the gas in the original subhalo has been consumed \citep{2014MNRAS.445.1056V}.
Multiple studies have argued that DCBH formation can occur shortly after the onset of Pop~III star formation, possibly at $z > 15$ \citep[e.g.,][]{2014MNRAS.445.1056V,Nabizadeh_2024}.

The expected DCBH number density is, however, highly sensitive to the critical LW flux. \citet{2014MNRAS.442.2036D} estimate a DCBH number density of $10^{-7}$–$10^{-12}\,\mathrm{cMpc^{-3}}$ at $z = 20$ for a critical LW flux of $J_\mathrm{crit} = 300$, in units of $10^{-21}\,\mathrm{erg\,s^{-1}\,cm^{-2}\,Hz^{-1}\,sr^{-1}}$. For a lower critical flux of $J_\mathrm{crit} = 30$, the number density increases to $\sim10^{-5}\,\mathrm{cMpc^{-3}}$ at the same redshift. Based on these studies, we adopt $z = 20$ as a plausible initial redshift at which at least one DCBH seed can exist within the survey area. Similar to the stellar remnant case (Sec.~\ref{meth:LS}), this choice also implies that atomic cooling halos $(\sim10^8$ M$_\odot$), the expected DCBH hosts, are common within the survey volume, as not every atomic cooling halo will form DCBHs. Using GUREFT, we expect $\sim0.02$ Mpc$^{-3}$ dex$^{-1}$ atomic cooling halos at $z\sim20$, while the estimated survey volume is $\sim8\times10^6$ Mpc$^{3}$ at $z\sim20$. 
However, we note that the value of the critical LW flux remains uncertain, and that DCBH formation is also influenced by additional environmental factors such as galactic winds, the escape fraction of LW photons \citep{2014MNRAS.442.2036D}, and the amplitude of baryon-dark matter streaming \citep{Schauer_2017}.

We adopt an initial DCBH mass of $10^{5}\,M_\odot$ \citep[e.g.,][]{2003ApJ...596...34B, 2010MNRAS.402.1249S,Becerra2018}, and assume continuous accretion at the Eddington rate ($\lambda_{\rm Edd}=1$) as an optimistic growth prescription.
Like the LS case, we obtain the maximum black hole mass by tracing the growth of a DCBH seed formed under the adopted initial conditions. Further details are presented in Appendix~\ref{app:growth}. 


\section{Results and Discussion} \label{sec:Results}

\subsection{Minimum Mass Requirement for Detection} \label{res:min_M}

\begin{figure*}[htbp]
\centering

\subfloat[\label{fig:MM_range}]{
    \includegraphics[width=0.415\linewidth]{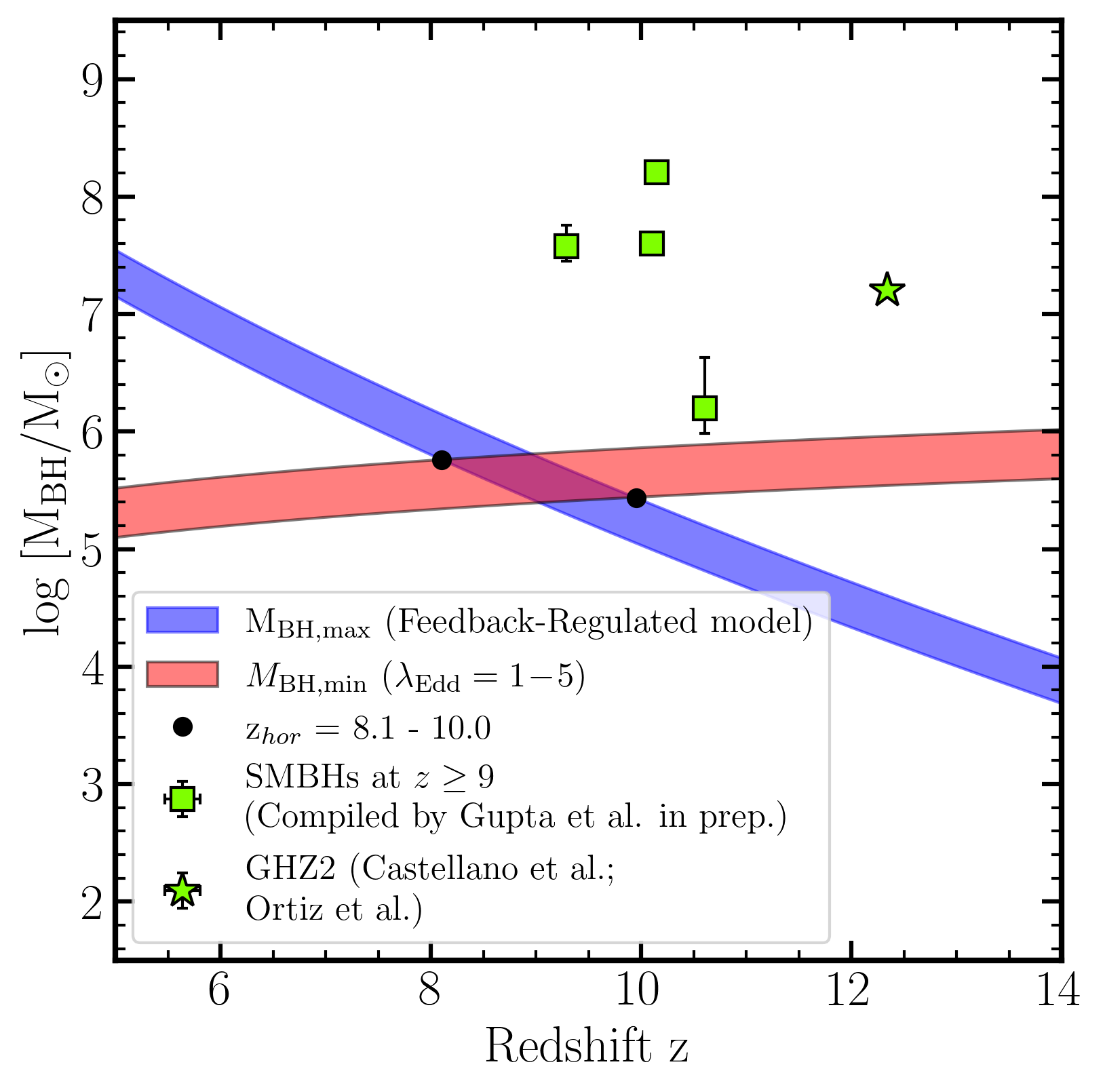}
}
\hfill
\subfloat[\label{fig:Density}]{
    \includegraphics[width=0.5\linewidth]{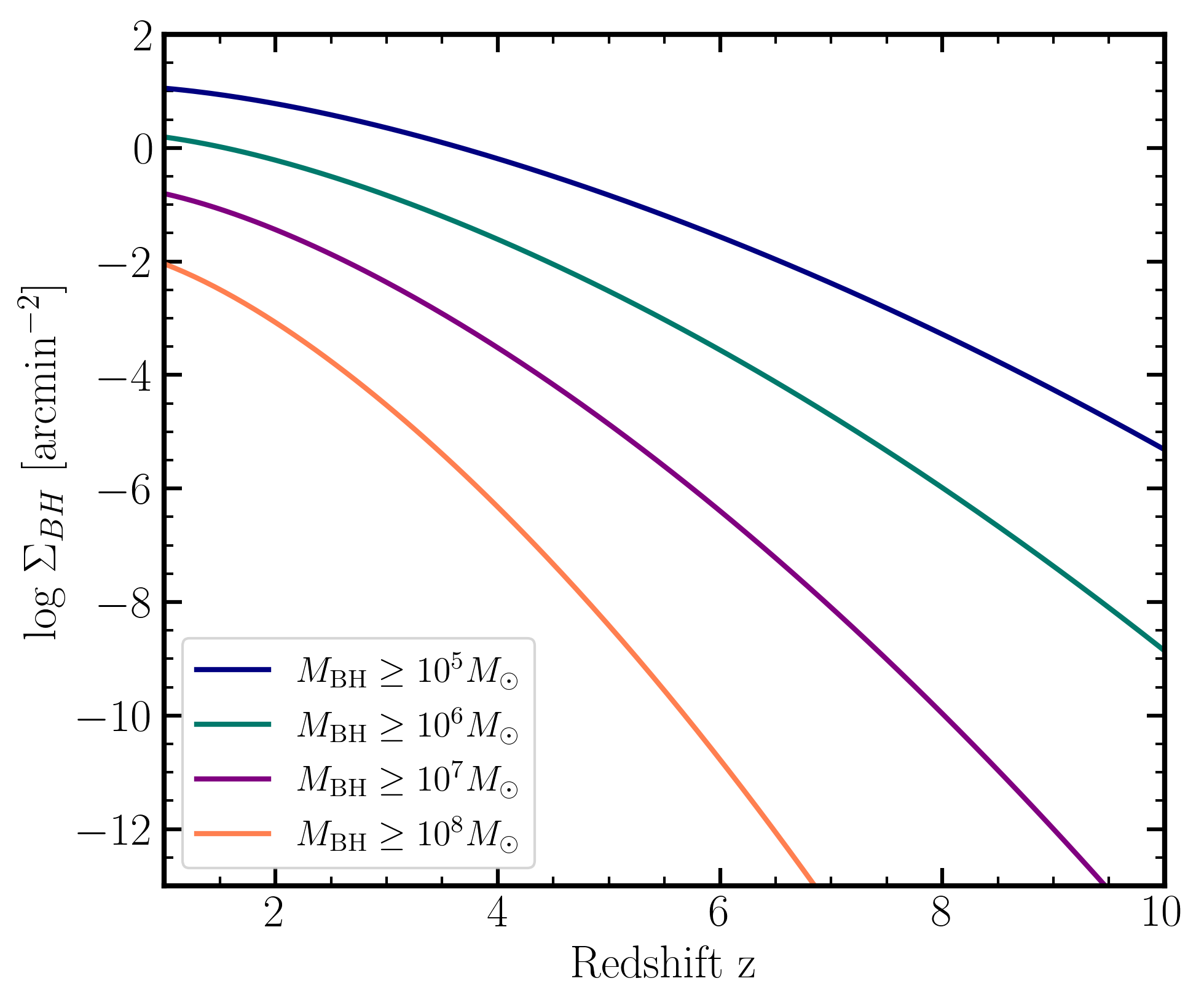}
}

\caption{
Summary of the feedback-regulated scenario.
\textit{Left:} Maximum and minimum black hole masses implied by abundance and detectability constraints, respectively. 
The blue shaded region shows the maximum black hole mass, such that their expected number within a fiducial survey volume of $10^{6}$--$10^{7}\,\mathrm{cMpc^3}$ is of order unity. 
The red shaded region shows the minimum detectable mass for a JWST NIRCam sensitivity of $1\,\mathrm{nJy}$ over the range $\lambda_{\mathrm{Edd}}=1$--$5$. 
Light green symbols with error bars denote reported JWST SMBH candidates at $z\gtrsim 9$. 
\textit{Right:} Surface number density of black holes above several mass thresholds as a function of redshift in the same scenario. 
The SMBH abundance rapidly declines toward high redshift, implying that detections of $\gtrsim10^{5}\,M_\odot$ black holes at $z\gtrsim10$ would require very large survey areas. 
Taken together, the two panels show that even under optimistic assumptions, feedback-regulated growth alone struggles to explain the highest-redshift SMBH candidates.
}
\label{fig:combined}
\end{figure*}

We use the methodology described above to compute the minimum SMBH mass required to produce observed flux densities above the detection threshold. The red shaded region in Figure~\ref{fig:MM_range} and the black band in Figure~\ref{fig:final} show this minimum mass as a function of redshift where we express the uncertainty in our modeling of the bolometric luminosity with an Eddington factor, such that $L_{\rm bol}\simeq \lambda_{\rm Edd} L_{\rm Edd}$. The red shaded region spans
$\lambda_{\rm Edd}=1-5$, which we adopt for a later simplified calculation. The black band in Figure~\ref{fig:final} spans $\lambda_{\rm Edd}=1-2$, which we use to estimate the quantity in Section~\ref{res:FR}.

%

Regardless of the formation or growth scenario, this minimum mass requirement must be satisfied for detection by the instrument. Having established the minimum mass required for detectability, we now consider the maximum mass allowed by SMBH abundance constraints across different formation and growth scenarios.

\subsection{Feedback-Regulated Model} \label{res:FR}

\begin{figure*}[htbp]
\centering
\includegraphics[width=\linewidth]{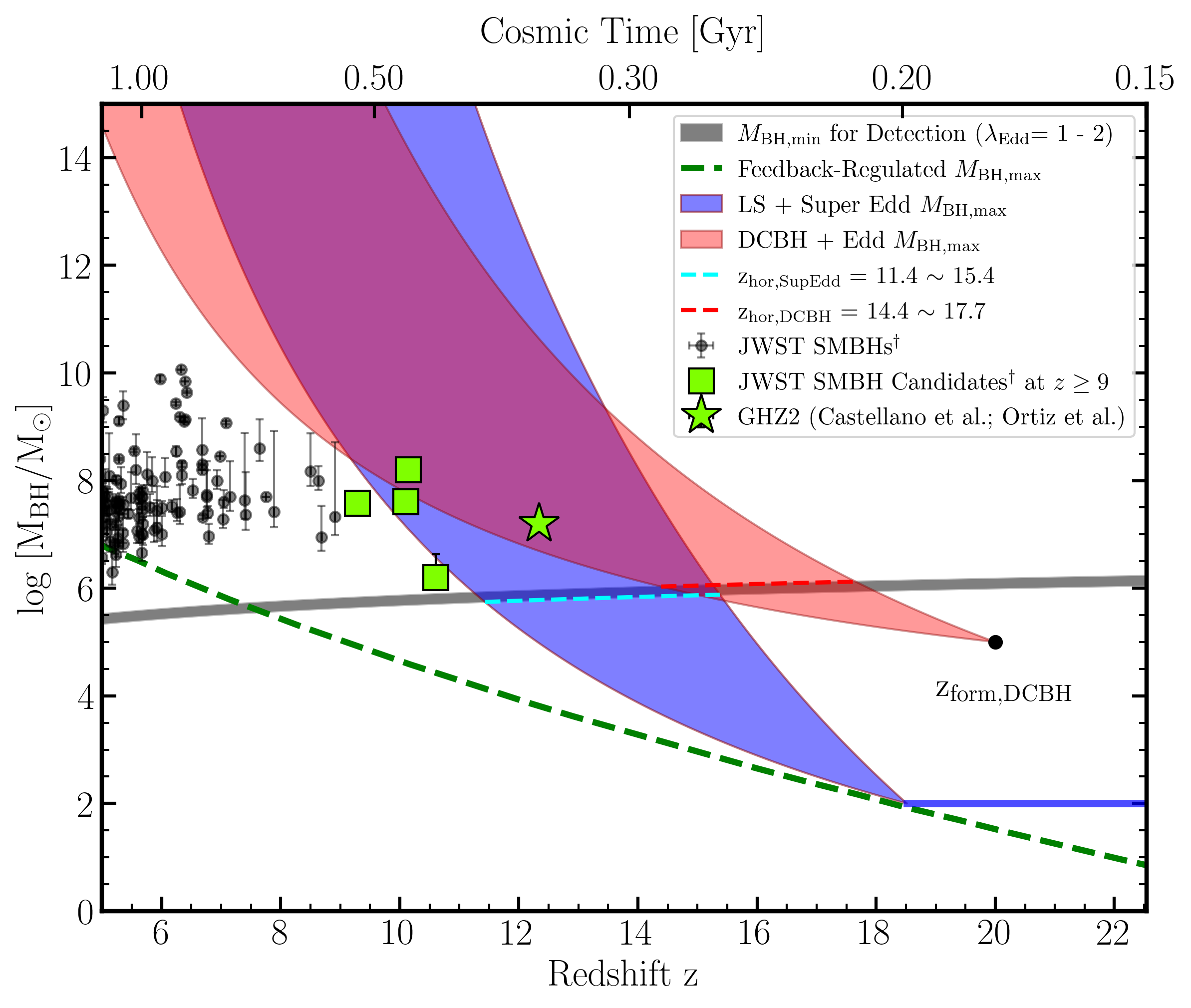} 
\caption{Redshift horizons for SMBH detectability with JWST/NIRCam photometry. The black solid line denotes the minimum black hole mass required for detection at a sensitivity of $1\,\mathrm{nJy}$. This curve corresponds to a band bounded by $\lambda_\mathrm{Edd}=1$ (upper) and $\lambda_\mathrm{Edd}=2$ (lower). The green dashed line represents the maximum black hole mass allowed by abundance constraints in the feedback-regulated model within a survey area of $10{,}000\,\mathrm{arcmin^2}$. The blue shaded region shows the growth tracks of LS black holes, assuming formation at $z=30$, a fallback time of $100\,\mathrm{Myr}$, and radiative efficiencies in the range $0.03$–$0.1$ (the initial pre-accretion phase is not shown). The red shaded region shows the growth of DCBHs, assumed to form at $z=20$ and accrete at the Eddington limit with the same range of radiative efficiencies. The intersections between the minimum mass curve and the LS and DCBH tracks, shown as cyan and red dashed lines, define the redshift horizons for each scenario. Black circles with error bars represent current SMBH observations compiled from numerous papers, while light green markers indicate SMBH candidates observed by JWST at $z\geq9$, including GHZ2 \citep[e.g.,][]{2023A&A...677A..88B, 2025ApJ...986..126K, 2025ApJ...989L...7T, ortiz2025significantevidenceagncontribution, 2026MNRAS.546ag086J}.
Under our optimistic LS and DCBH prescription, the redshift horizon at which JWST can photometrically detect SMBHs may extend to $\sim17$ exceeding our current empirical limit. However, whether JWST can spectroscopically confirm SMBHs at these redshifts remains uncertain. We assess this in  Section \ref{res:spec}.}
\label{fig:final}
\end{figure*}

For an approximate estimate of the maximum SMBH mass detectable in the feedback-regulated model, we consider a fiducial survey volume of $10^{6}$–$10^{7}\,\mathrm{cMpc^{3}}$ and multiply with the SMBH host halo number density predicted by the Press–Schechter formalism at each redshift. 
Figure~\ref{fig:MM_range} illustrates the resulting maximum mass constraint under this simplified approach, where the blue shaded region represents the constraint that the expected number of SMBHs within the survey volume is of order unity. The red shaded region indicates the minimum detectable mass for a fixed JWST NIRCam sensitivity of $1\,\mathrm{nJy}$, over a range of luminosities (corresponding to $1$–$5$ times the fiducial Eddington value). For comparison, we reproduce select SMBH candidates observed by JWST at $z\gtrsim9$ ({\it ligh-green symbols}). We note that the lower boundary here, corresponding to $L_{\rm bol}\simeq 5 L_{\rm Edd}$, may be overly optimistic, given that local AGN are observed to predominantly emit at sub-Eddington or near-Eddington luminosities \citep[e.g.,][]{2013ARA&A..51..511K, 2014ARA&A..52..589H}.

The intersection of the two regions constrains the horizon to $z \sim 8-10$.  Notably, even for these optimistic assumptions, current observations of high-redshift SMBH candidates already lie beyond the predicted horizon. 
A more detailed calculation, based on Equation~(\ref{N_mh}), yields a horizon at $z\sim7.3$, corresponding to the intersection between the minimum mass curve ({\it black solid line}) and the corresponding maximum mass curve ({\it dark green dashed line}) in Figure~\ref{fig:final}, in even more serious tension with the observed high-$z$ AGN.



Evaluating Equation~(\ref{N_mh}), we derive the SMBH number density for different masses (Fig.~\ref{fig:Density}). Within a survey area of $10,000~\mathrm{arcmin}^2$, a single $10^5 \,M_\odot$ black hole is expected to be found at $z\lesssim9$. For SMBH candidates such as UHZ1 and GHZ2, detected at $z\gtrsim 10$ with inferred masses of $\sim10^7\,M_\odot$ \citep[e.g.,][]{2024ApJ...960L...1N, ortiz2025significantevidenceagncontribution}, their surface number densities would be even lower. Our results are therefore consistent with previous studies suggesting that feedback-regulated growth alone cannot fully account for the rapid formation of $\sim10^9\,M_\odot$ black holes within the first billion years of cosmic history \citep[e.g.,][]{2004ApJ...614L..25Y, 2005ApJ...633..624V, 2009ApJ...701L.133A}. Instead, SMBHs in the early universe may form in overmassive configurations, compared to the virial and stellar mass of the host system. Any approach to local $M-\sigma$ like relations, arising from feedback self-regulation, would thus occur only on longer timescales \citep[e.g.,][]{Hu_Normalcy_2025,Zhang_Normalcy_2026}.



\subsection{Horizon for the Light Seed Model} \label{res:LS}
Under the LS super-Eddington accretion scenario, the growth of light seed black holes is characterized by continuous accretion at twice the Eddington limit following a fallback time of $100\,\mathrm{Myr}$, representing the evolution of the maximum mass. The blue shaded region in Figure \ref{fig:final} shows the corresponding growth tracks of LS black holes starting from an initial mass of $100\,M_\odot$, with radiative efficiency in the range $0.03$–$0.1$. The redshift horizon in this scenario is $z\sim11.5-15.5$, defined by the intersection with the minimum mass curve. 
This result represents an optimistic upper bound on the JWST/NIRCam redshift horizon for this scenario, and is consistent with the current observational frontier, including UHZ1, GN-z11, and GHZ2 \citep[e.g.,][]{2024ApJ...960L...1N, 2024Natur.627...59M, 2023A&A...677A..88B, 2024ApJ...972..143C, ortiz2025significantevidenceagncontribution}.

\subsection{Horizon for the Heavy Seed Model} \label{res:DCBH}
In contrast to the LS scenario, DCBHs reach the minimum mass required for detection earlier, within $\sim 100\,\mathrm{Myr}$ after seed formation, given their larger initial mass. In Figure~\ref{fig:final}, the red shaded region represents black holes that form as DCBHs and subsequently undergo Eddington-limited accretion. The resulting redshift horizon, again determined by the intersection with the detection mass threshold, is $z \sim 14.5-18$. 
Although this scenario involves a lower accretion rate compared to the light seed case with super-Eddington accretion, the much larger initial seed mass results in a significantly higher redshift horizon. This difference in redshift horizons may provide a potential means of distinguishing between the two seed scenarios, subject to uncertainties in the underlying assumptions. In particular, detecting an SMBH beyond $z\sim16$ would favor a heavy seed origin over the LS alternative.

Unlike the feedback regulated scenario, both the light and heavy seeding models are consistent with the SMBH candidates detected and confirmed by JWST to date  \citep[e.g.,][]{2023Natur.621...51D, 2023ApJ...959...39H, 2023ApJ...954L...4K, 2023ApJ...957L...7K, 2023ApJ...953L..29L, 2023ApJ...953..180S, 2023A&A...677A.145U, 2024NatAs...8..126B, 2024Natur.628...57F, 2024ApJ...964...39G, 2024A&A...685A.121L, 2024ApJ...963..129M, 2024ApJ...964...90S, 2024MNRAS.531..355U, 2024ApJ...969L..13W, 2024ApJ...966..176Y, 2025ApJ...980L..29A, 2025ApJ...986..177B, 2025ApJ...986..126K, 2025ApJ...989...75N, 2025NatCo..16.9830T, 2026MNRAS.546ag086J, 2026ApJ...996...93L, 2026Natur.656..329N}. The substantial redshift gap of $\Delta z \sim 3-5$ between GHZ2 and the predicted horizons suggests the exciting possibility to discover additional SMBHs beyond the currently observed redshift frontier.

\subsection{Prospects for the Roman Space Telescope} \label{res:RST}
We also note that similar calculations can be carried out for the Nancy Grace Roman Space Telescope (RST), which has been launched in August 2026. Although JWST's NIRCam is $\sim3\times$ more sensitive in the F150W filter ($1.33$–$1.67\,\mathrm{\mu m}$) than RST in its similar F158 filter ($1.36$–$1.77\,\mathrm{\mu m}$), RST offers a field of view that is a factor of $\sim100$ greater than that of NIRCam. As a result, JWST is better suited for detecting lower mass black holes at higher redshifts, while RST will be more efficient at capturing rarer, more massive black holes over a wide area. We therefore expect these two facilities to play complementary roles in studies of the first black holes, jointly enabling empirical investigations of early black hole formation and growth.

\subsection{Spectroscopic Confirmation} \label{res:spec}
Although we have presented the redshift horizons for JWST photometric detection of black holes, spectroscopic confirmation is ultimately required to establish that a detected source is indeed powered by a black hole. We here therefore briefly explore feasible strategies for confirming black holes with masses of $10^{6}$–$10^{8}\,M_\odot$ at $z\sim16$, corresponding to the redshift and mass range identified in our analysis.

Broad-line diagnostics using $H\alpha$ and $H\beta$, which are commonly regarded as the most secure methods for SMBH confirmation, are not applicable at these redshifts. At $z \gtrsim 10$, both Balmer lines are shifted beyond the wavelength coverage of NIRSpec. Although MIRI MRS extends to longer wavelengths, spectroscopic confirmation using these lines would require extremely long exposure times.
Adopting the $H\alpha$ line luminosities of the GLIMPSE intermediate-mass black hole sources with masses of $10^{6}$–$10^{7}\,M_\odot$, the corresponding line flux becomes $\sim2\times10^{-19}\,\mathrm{erg\,s^{-1}\,cm^{-2}}$ at $z=15.5$. In this case, MIRI MRS would require approximately $100$ days of exposure to achieve a peak $\mathrm{SNR}\sim3$, indicating that spectroscopic confirmation with MIRI MRS would not be feasible \citep{2025arXiv250920452F}. We therefore focus on alternative spectroscopic methods that remain accessible with NIRSpec.

A useful reference case is provided by the NIRSpec analysis of GN-z11, where the NIV] doublet and NIII] multiplet were used to infer gas densities that are difficult to explain via photoionization by typical star-forming galaxies. The reported target has $\log M_\mathrm{BH} = 6.2$ at $z = 10.6$ \citep{2024Natur.627...59M}. Under background-limited scaling, if a similar object was located at $z = 16$, reproducing a comparable line-based analysis would require approximately $\sim8$ times longer exposure time. This implies that NIRSpec/G235M with F170LP would require $\sim29~\mathrm{hrs}$, while the prism/CLEAR mode would require $\sim57.3~\mathrm{hrs}$ to detect NIV] $\lambda1486$ and NIII] $\lambda1750$ at comparable significance \citep{2023A&A...677A..88B}.

In addition, UV emission lines such as N{\sc{V}} $\lambda1240$ can serve as important SMBH diagnostics, as they require a hard ionizing radiation field. As a concrete example, we adopt the N{\sc V} $\lambda1240$ line luminosity of the JADES-NS-GS00021842 narrow-line AGN target, which is expected to host a black hole mass comparable to those considered here for Eddington ratios of $0.01$–$2$ \citep{2025A&A...697A.175S}. Shifting its intrinsic N{\sc V}~$\lambda1240$ luminosity from the source redshift $z = 7.98$ to $z \sim 15.5$–$17$, we find that with NIRSpec G235M/F170LP, an exposure time of $\sim100~\mathrm{hr}$ yields a peak N{\sc V} $\lambda1240$ of $\mathrm{SNR} \sim 3\sigma$. Although the SNR depends primarily on the intrinsic line luminosity, if an object similar to JADES-NS-GS00021842 exists near our predicted redshift horizon, the detection of N{\sc V}~$\lambda1240$ would provide a viable path toward spectroscopic confirmation of an SMBH.

Returning to our main conclusion, if SMBHs at $z\sim16$ can not only be detected but also confirmed spectroscopically using the approaches outlined above, such observations may offer a complementary way to assess which seeding scenario is more plausible, probing a regime that is difficult to reconcile even with optimistic light seed growth scenarios.



\section{Conclusions} \label{sec:Conclusion}

With the advent of JWST, our observational horizon has rapidly expanded, opening a new window into the early universe. These unprecedented observations provide a unique opportunity to probe the formation and early growth of SMBHs through empirical evidence.

In this work, we investigate the maximum redshift at which SMBHs can be observed with JWST under several representative seeding and growth scenarios. Although the detailed growth of SMBHs is expected to depend strongly on their environments, the predicted redshift horizons suggest that JWST may be capable of detecting SMBHs at even earlier cosmic epochs than currently observed. We further explore whether differences between these horizons could distinguish between SMBH seeding scenarios, taking into account current modeling uncertainties. Finally, we assessed the feasibility of spectroscopic confirmation for such distant sources. The main findings of this study are summarized as follows:

\begin{itemize}
\item For the feedback-regulated scenario, we obtain a redshift horizon of $z\sim7$. This value is already exceeded by several SMBHs observed by JWST, implying that the feedback-regulated model alone is insufficient to explain the early evolution of SMBHs. 

\item For the LS model with super-Eddington accretion, we obtain a redshift horizon of $z\sim15$, compared to $z\sim17-18$ for the DCBH model with Eddington-limited accretion. Both extend beyond the current highest-redshift SMBH candidate at $z = 12.3$ \citep[GHZ2; e.g.,][]{ortiz2025significantevidenceagncontribution}. This suggests promising prospects for detecting SMBHs at even earlier cosmic epochs. 

\item The gap between the redshift horizons for the LS and DCBH scenarios may offer another way to distinguish between the two seeding models, although this possibility remains subject to uncertainties in the underlying assumptions. According to our analysis, the detection of an SMBH at $z\gtrsim16$ would favor a heavy seed origin.

\item Finally, we examine whether such distant sources could be confirmed spectroscopically. This may be achieved through NIRSpec line-based diagnostics similar to those applied to GN-z11 \citep[e.g.,][]{2023A&A...677A..88B}, requiring $\sim8$ times longer exposure, or through detection of strong rest-frame UV AGN signatures (e.g., N{\sc V}) with an exposure time of $\sim100\,\mathrm{hr}$. 

\end{itemize}

These results suggest that JWST can significantly extend the observational frontier of SMBHs toward earlier cosmic epochs. In addition, the upcoming Roman Space Telescope (RST) may play a complementary role in identifying rarer but more massive SMBHs owing to its much larger field of view. Such observations may play a transformative role in understanding the seeding and early growth of SMBHs, thus probing the elusive first steps in cosmological structure formation.


\begin{acknowledgments}
The authors thank Ansh Gupta for sharing his compilation of JWST-observed SMBHs reported in the literature. AT acknowledges support from UT Austin's College of Natural Sciences.
VB acknowledges support from the Josey Centennial Professorship in Astronomy at UT Austin. ChatGPT (OpenAI) was used to assist with grammar correction and language editing during manuscript preparation. HS reviewed the suggested edits.
\end{acknowledgments}





%

\vspace{5mm}

\software{astropy \citep{2013A&A...558A..33A,2018AJ....156..123A},  
          Colossus \citep{2018ApJS..239...35D}, GUREFT \citep{Yung2024}
}


\appendix
\section{Seed Growth Curves for the LS and DCBH Abundance Constraints}
\label{app:growth}
The abundance constraints for the LS and DCBH scenarios are obtained by tracing the growth of representative seed black holes under the setups described in Sections~\ref{meth:LS} and \ref{meth:DCBH}. We assume that each seed undergoes continuous isotropic Eddington or super-Eddington accretion ($\lambda_\mathrm{Edd}=2$), with a rate given by
\begin{equation}
\label{App: Edd_rate}
    \dot{M}_\mathrm{Edd} = \lambda_\mathrm{Edd}\frac{L_\mathrm{Edd}}{\eta\,c^2} = \lambda_\mathrm{Edd}\frac{4\pi G m_\mathrm{p}}{\eta\,c\,\sigma_\mathrm{T}}M_\mathrm{BH},
\end{equation}
where $\sigma_\mathrm{T}$ is the Thomson scattering cross-section, $m_p$ the mass of a proton, and $\eta$ the radiative efficiency (with a range of $0.03$–$0.1$). Rewriting Equation~\ref{App: Edd_rate} in terms of redshift gives
\begin{equation}
\label{App: M_DE}
    \frac{dM_\mathrm{BH}}{dz} = -\frac{\lambda_\mathrm{Edd}}{(1+z)H(z)}\frac{4\pi G  m_\mathrm{p}}{\eta\, c\, \sigma_\mathrm{T}} M_{\mathrm{BH}},
\end{equation}
where
\begin{equation}
\label{App: Hz}
    H(z) = H_0\left[ \Omega_m(1+z)^3 +\Omega_{\Lambda} \right]^{1/2}\mbox{\ .}
\end{equation}

The upper mass limit at each redshift is then obtained by solving Equation~\ref{App: M_DE} for the adopted seed models. As discussed in Sections~\ref{meth:LS} and \ref{meth:DCBH}, we apply a $100\,\mathrm{Myr}$ fallback time after the initial redshift of $z=30$ in the LS scenario before evolving Equation~\ref{App: M_DE}. For the DCBH scenario, we instead adopt an initial redshift of $z=20$ without a fallback delay.


\bibliography{Main, sample7}
\bibliographystyle{aasjournal}

\end{document}